\documentclass[journal]{IEEEtran}
\usepackage[numbers,sort&compress]{natbib}

\ifCLASSINFOpdf
   \usepackage[pdftex]{graphicx}
\else
\fi
\usepackage{xcolor}

\usepackage{amssymb}
\usepackage{amsmath}
\usepackage{framed,multirow}
\usepackage{array}
\usepackage{url}

\begin{document}
	
%
\title{Efficient 3D Fully Convolutional Networks for Pulmonary Lobe Segmentation in CT Images}
%
%
%

\author{Hoileong Lee, Tahreema Matin, Fergus Gleeson, and Vicente Grau \vspace{-8mm}
\thanks{This work was supported by the CRUK and EPSRC Cancer Imaging Centre in Oxford. The work of H. Lee was supported by a fellowship co-funded by the Malaysian Ministry of Higher Education and Universiti Malaysia Perlis. (\textit{Corresponding author: Hoileong Lee.})}
\thanks{H. Lee is with the Institute of Biomedical Engineering, Department of Engineering Science, University of Oxford, Oxford OX3 7DQ, U.K., and also with the School of Mechatronic Engineering, Universiti Malaysia Perlis, 02600 Arau, Perlis, Malaysia (e-mail: hoileong.lee@eng.ox.ac.uk).}
\thanks{T. Matin and F. Gleeson are with the Department of Radiology, Churchill Hospital, Oxford OX3 7LE, U.K.}
\thanks{V. Grau is with the Institute of Biomedical Engineering, Department of Engineering Science, University of Oxford, Oxford OX3 7DQ, U.K.}
}

%
%

\markboth{}%
{Shell \MakeLowercase{\textit{et al.}}: Bare Demo of IEEEtran.cls for IEEE Journals}
%



\maketitle

\begin{abstract}
The human lung is a complex respiratory organ, consisting of five distinct anatomic compartments called lobes. Accurate and automatic segmentation of these pulmonary lobes from computed tomography (CT) images is of clinical importance for lung disease assessment and treatment planning. However, this task is challenging due to ambiguous lobar boundaries, anatomical variations and pathological deformations. In this paper, we propose a high-resolution and efficient 3D fully convolutional network to automatically segment the lobes. We refer to the network as Pulmonary Lobe Segmentation Network (PLS-Net), which is designed to efficiently exploit 3D spatial and contextual information from high-resolution volumetric CT images for effective volume-to-volume learning and inference. The PLS-Net is based on an asymmetric encoder-decoder architecture with three novel components: (i) 3D depthwise separable convolutions to improve the network efficiency by factorising each regular 3D convolution into two simpler operations; (ii) dilated residual dense blocks to efficiently expand the receptive field of the network and aggregate multi-scale contextual information for segmentation; and (iii) input reinforcement at each downsampled resolution to compensate for the loss of spatial information due to convolutional and downsampling operations. We evaluated the proposed PLS-Net on a multi-institutional dataset that consists of 210 CT images acquired from patients with a wide range of lung abnormalities. Experimental results show that our PLS-Net achieves state-of-the-art performance with better computational efficiency. Further experiments confirm the effectiveness of each novel component of the PLS-Net.
\end{abstract}

\vspace{-3mm}

%
\IEEEpeerreviewmaketitle

\section{Introduction}
%
%
%
%
\IEEEPARstart{A}{natomically}, the human lungs are subdivided into lobes by pulmonary fissures. The right major and minor fissures separate the right lung into three lobes (upper, middle and lower), whereas the left major fissure separates the left lung into two lobes (upper and lower). Each lobe houses its own subtrees of airways and blood vessels, and is usually considered as an independent functional unit. Lobe-wise analysis is of significant interest since many pulmonary diseases are known to affect the lung lobes differently, selectively and in complex progression patterns. For example, centrilobular emphysema, cystic fibrosis and tuberculosis predominantly affect the upper lobes \cite{Nemec:2013aa}, while interstitial pneumonia and panlobular emphysema mainly involve the lower lobes \cite{Nemec_2013}. The early onset and heterogeneity of these diseases are poorly captured in global lung function indices from pulmonary function tests, particularly when there is functional compensation from healthy lung tissues \cite{Stahr2016}.

The non-uniformity of pulmonary lobe involvement in pathology makes \textit{in vivo} imaging assessment of the lungs highly desirable. Computed tomography (CT), with its high spatial resolution and excellent anatomic detail, has long been considered as the modality of choice for assessing the lung morphology. Segmentation of pulmonary lobes from chest CT images allows a more localised, sensitive diagnosis of lung diseases and more accurate treatment planning and monitoring. For instance, CT lobar volumetry and densitometry are performed to quantify spatial distribution and severity of emphysema, identifying patients with heterogeneous emphysema for lung volume reduction surgeries \cite{Ostridge2016}. CT-based lobe segmentation is also required to derive lobar function measurements from functional imaging, such as hyperpolarised gas magnetic resonance imaging, which yields images with no visible interlobar fissures \cite{Matin2017,Lee2017}. It also enables the study of lobar structure-function relationships in different lung diseases, improving understanding of pathophysiology.

Although manual lobe delineation is possible, it is laborious, inefficient, and prone to intra- and inter-observer variability. A fast, accurate and automatic method for lung lobe segmentation is therefore highly demanded in clinical practice. The method also has to be reproducible and handle pathological cases. However, automating the segmentation is a challenging task due to complicating factors such as anatomical variations, pathological deformations, missing or incomplete fissures, and imaging artefacts. The presence of nodules, atelectasis, fibrosis, emphysema, accessory fissures or other fissure-like structures introduces additional problems. \vspace{-3mm}

\subsection{Related work}

Many automatic and semi-automatic methods for lung lobe segmentation in CT images have been proposed over the past decades \cite{Doel2015}. Since pulmonary fissures define the interlobar boundaries, early works formulated lobar segmentation as a fissure detection problem. Approaches ranging from atlas registration \cite{LiZhang2006} to supervised learning \cite{VanRikxoort2009, Schmidt-Richberg2012} and a combination of local filtering with curve fitting \cite{JiantaoPu2009, Gu2012, Ross2013} were proposed to automatically delineate the visible fissures and interpolate incomplete ones. These methods, however, are not robust to cases with incomplete fissures and severe pathologies.

Instead of solely relying on fissure information, \citet{Kuhnigk2005} incorporated auxiliary anatomical cues from pulmonary vessels to segment the lobes with a 3D watershed algorithm, using the anatomical trait of lobes having separate bronchial and vascular trees. This shift from fissure- to anatomy-guided lobar segmentation can be seen in later work such as \citet{Ukil2009}, \citet{Doel2012} and \citet{Lassen2013}, as well as more recently by \citet{Bragman2017} and \citet{Giuliani2018}. For example, \citet{Lassen2013} extended the framework in \cite{Kuhnigk2005} by including information from automatic segmentations of the lungs, fissures, airways and vessels into the cost image for watershed segmentation, while \citet{Bragman2017} incorporated a group-wise fissure prior in addition to this anatomical information. Even though these methods have, to some extent, been successful in practice, they still suffer from several limitations: (i) high computational time since a complex, multi-stage pipeline is involved; (ii) the segmentation performance is dependent on the quality of the atlas and/or automatic segmentations of pulmonary structures; and (iii) traditional automatic methods  based on hand-crafted features tend to depend on multiple parameters and struggle with large variations of appearance and shape, leading to poor performance in diseased and abnormal anatomy cases.

The recent success of deep learning, particularly fully convolutional networks (FCNs) \cite{Shelhamer2017}, has opened up new possibilities for automatic medical image segmentation. The main advantage of FCNs over traditional segmentation methods is that the features are automatically and hierarchically learned from the data to optimise the objective function in an end-to-end manner, leading them to achieve state-of-the-art performance in many segmentation tasks. \citet{Harrison2017} proposed a 2D deeply supervised FCN with multi-path connections, namely progressive holistically-nested network (P-HNN), for segmenting the lungs in CT images. Later, \citet{George2017} used the 2D P-HNN coupled with a 3D random walker algorithm (P-HNN+RW) to delineate the lung lobes in 3D. This 2D FCN-based method processes each CT slice independently and ignores 3D context, which is arguably a suboptimal use of 3D data and could affect segmentation accuracy, especially for cases with incomplete fissures in which anatomical information from neighbouring slices is usually needed. \looseness=-1

A straightforward way to encode 3D spatial information in volumetric CT images is to extend the convolution kernels in FCNs from 2D to 3D. Recently, \citet{Imran2018} introduced a 3D FCN architecture inspired by the P-HNN and dense V-network \cite{Gibson2018}, called progressive dense V-network (PDV-Net), for automated segmentation of pulmonary lobes. Similarly, \citet{Park2019} presented an automatic lung lobe segmentation method based on the 3D U-Net \cite{Ronneberger2016} architecture, whereas \citet{Ferreira2018} proposed a fully regularized V-Net (FRV-Net). Albeit promising results are reported in these studies, a number of limitations apply to the proposed 3D FCNs: (i) the networks have high computational complexity and large parameter count, which means that they need long training time and large datasets to effectively optimise millions of model parameters; (ii) the networks are memory-inefficient and cannot be trained with the entire high-resolution volumetric CT images even on a modern high-end graphics processing unit (GPU) (e.g. NVIDIA Geforce GTX 1080 Ti with 11 GB of memory). Due to these constraints, they are trained with either 3D patches \cite{Ferreira2018,Park2019} or aggressively downsampled 3D volumes \cite{Imran2018}, which would compromise the network capability to capture global context or to learn high-resolution 3D feature representations and thus degrade the segmentation performance; and (iii) the networks have a relatively small receptive field due to their limited depth constrained by the GPU memory, which implies that the spatial extent from which the feature information can be extracted is restricted. For example, the neurons in even the deepest layer of the 3D U-Net in \cite{Ronneberger2016} only have a receptive field of $68^{3}$ voxels (the effective receptive field size is much smaller \cite{Luo2016,Liu2015}). This limits the integration of global context, which has been shown to be essential for accurate fissure detection \cite{Gerard2018}.\vspace{-3mm}

\subsection{Contributions}

To address the aforementioned limitations, we present a novel automated lung lobe segmentation method based on a high-resolution, efficient 3D FCN which we refer to as Pulmonary Lobe Segmentation Network (PLS-Net). In contrast to recent FCN-based approaches that process a volumetric CT image in a slice-wise, 3D patch-wise or downsampled volume-wise manner, our PLS-Net is designed to work with the whole high-resolution CT volume for effective and efficient volume-to-volume learning and inference. Specifically, the network takes a preprocessed CT volume as input and outputs correspondingly-sized volumetric lobar segmentations in an end-to-end fashion by leveraging an asymmetric encoder-decoder architecture with 3D depthwise separable convolutions, dilated residual dense blocks, and input reinforcement. The main contributions of this work are summarised below:

\begin{itemize}
	\item We extend the depthwise separable (DS) convolution presented by \citet{Sifre2014} to 3D. 3D DS convolutions factorise each regular 3D convolution into two simpler operations, resulting in a 3D model that is light-weight and computationally efficient. To the best of our knowledge, this is the first work that applies 3D DS convolutions for medical image segmentation.
	\item  We propose a dilated residual dense block (DRDB) to efficiently enlarge the receptive field of a network and capture wide-ranging, dense multi-scale context features for segmentation.
	\item We introduce an input reinforcement (IR) scheme to improve the segmentation performance by reinforcing input spatial information after each downsampling stage of the network.
	\item We formulate a high-resolution and efficient 3D FCN, named PLS-Net, based on 3D DS convolutions, DRDBs and IR for the task of pulmonary lobe segmentation.
	\item We validate the proposed PLS-Net on a diverse dataset comprising 210 lung CT scans from multiple institutions. Experimental results show that the PLS-Net can efficiently segment the lobes directly from high-resolution volumetric CT images  obtaining state-of-the-art results.\
	\end{itemize}\vspace{-2mm}

\section{Materials}\label{sec:materials}

The dataset used in this study consists of 210 chest CT images of different subjects, which  were collected retrospectively from the Lung Tissue Research Consortium (LTRC) (n = 100) \citep{Karwoski2008}, the Lung Image Database Consortium (LIDC-IDRI) (n = 80) \citep{Freymann2011}, and our in-house chronic obstructive pulmonary disease (COPD) clinical trial (n = 30) \citep{Matin2017}. These images were acquired at different clinical centres using a variety of scanners, reconstruction kernels and imaging protocols. The slice thickness of the scans ranged from 0.50 mm to 1.50 mm, and the in-plane resolution varied between 0.53 mm and 0.88 mm. The imaged cohorts included subjects with various lung diseases such as lung cancer, COPD and interstitial lung disease. Common lung parenchymal abnormalities including fibrosis, nodules, emphysema, ground-glass opacity, reticular opacity, honeycombing and bronchiectasis were observed across the dataset. 147 CT images were found to have at least one incomplete fissure{\footnote{An incomplete fissure was defined as the fissure with a visual integrity of less than 90\% on CT.} (most frequently the right minor fissure) while accessory fissures were detected in 19 scans.}

Reference segmentations of pulmonary lobar structures were used as the gold standard for training and evaluation. For the LTRC dataset, lobar segmentations were provided together with the original scans, which were delineated semi-automatically by a LTRC radiologist. Specifically, the radiologist first manually defined the lung lobes on a few slices, and the lobar segmentations were then obtained using shape-based interpolation and gradient edge information, followed by manual corrections if needed. The remaining CT images were segmented semi-automatically at our centre using the Pulmonary Toolkit (PTK) \citep{Doel2012} and ITK-SNAP \citep{Yushkevich2006}. Manual refinement was then performed by an experienced radiologist.\vspace{-2mm}

\section{Methods}

In this section, we first describe the core building blocks of our proposed PLS-Net: the 3D depthwise separable convolution and dilated residual dense block. We then present the complete network architecture and its implementation details.\vspace{-2mm}

\subsection{3D depthwise separable convolution}

\begin{figure}[!t]
    \setlength\abovecaptionskip{-0.1\baselineskip}
	\centering
	\includegraphics[scale=0.4]{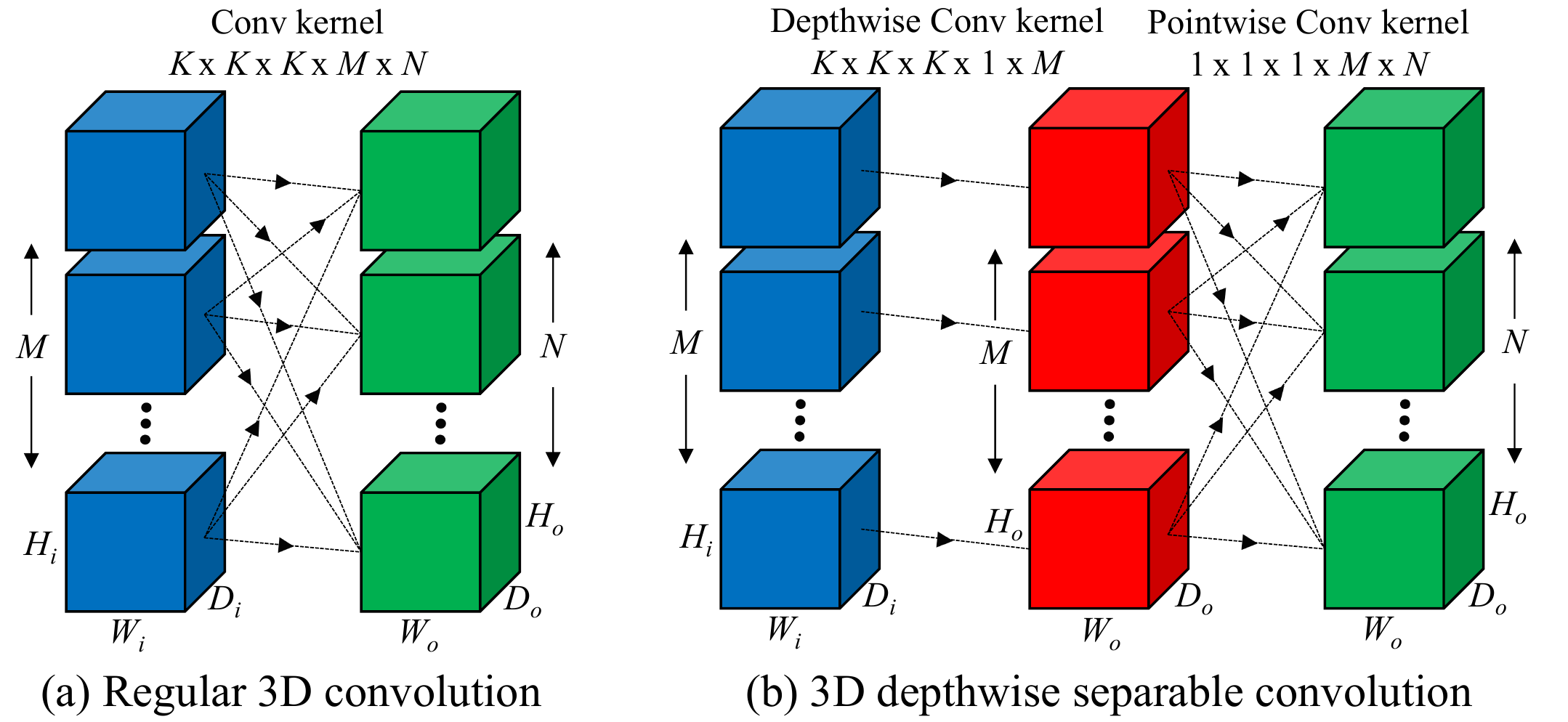}
	\caption{The difference between a regular 3D convolution and a 3D depthwise separable convolution.}
	\label{fig: depthwise}
	\vspace{-5mm}	
\end{figure}

Inspired by the MobileNet \cite{Howard2017} and Xception \cite{Chollet2017} architectures, our PLS-Net uses 3D depthwise separable (DS) convolutions, the 3D extension of the original 2D DS convolutions \cite{Sifre2014}. As shown in Figure \ref{fig: depthwise}, a 3D DS convolution is a factorisation of a regular 3D convolution into a depthwise convolution and a pointwise convolution. The depthwise convolution performs a 3D spatial convolution independently over each input channel, while the pointwise convolution applies a $1 \times 1 \times 1$ convolution to combine the depthwise convolution outputs and project them onto a new channel space. 

In essence, a regular 3D convolution (Figure \ref{fig: depthwise}(a)) learns both spatial (height, width and depth) and cross-channel correlations simultaneously with a 5D convolution kernel tensor $W\in\mathbb{R}^{K\times K \times K \times M \times N}$, where $K$ is the spatial dimension of the kernel (for simplicity we assume a cube-shaped kernel), $M$ is the number of input channels, and $N$ is the number of output channels.  It maps an input tensor $X\in\mathbb{R}^{H_i\times W_i \times D_i \times M}$ to an output tensor $Y\in\mathbb{R}^{H_o\times W_o \times D_o \times N}$, where $H$, $W$ and $D$ represent the height, width and depth of the feature map in the tensor, using the following linear mapping:
\begin{equation}\label{eq:1}
\displaystyle Y_{h,w,d,n} = \sum\limits_{i,j,k,m} X_{h+i,w+j,d+k,m} \cdot W_{i,j,k,m,n}
\end{equation}
The 3D convolution operation comes with high computational costs and large number of parameters. DS convolutions make this operation more efficient by explicitly factoring it into two simpler steps to capture spatial and cross-channel correlations separately. In a 3D DS convolution (Figure \ref{fig: depthwise}(b)), the 5D convolution kernel tensor $W$ of a regular 3D convolution is factorised into a depthwise convolution kernel tensor $D\in\mathbb{R}^{K\times K \times K \times M}$ and a pointwise convolution kernel tensor $P\in\mathbb{R}^{M \times N}$. This factorisation can be expressed as:
\begin{equation}\label{eq:2}
\displaystyle W_{i,j,k,m,n} = D_{i,j,k,m,}\cdot P_{m,n} 
\end{equation}
Based on (\ref{eq:2}), the output tensor $Y$ can be computed from the input tensor $X$ via a depthwise convolution followed by a pointwise convolution:
\begin{equation}\label{eq:3}
 \displaystyle Z_{h,w,d,m} =  \sum\limits_{i,j,k}  X_{h+i,w+j,d+k,m} \cdot D_{i,j,k,m}
\end{equation}
\begin{equation}\label{eq:4}
\displaystyle Y_{h,w,d,n} =  \sum\limits_{m} Z_{h,w,d,m} \cdot P_{m,n}
\end{equation}
where $Z\in\mathbb{R}^{H_o\times W_o \times D_o \times M}$. Note that the depthwise convolution can be performed with different dilation rates \cite{Chen2018} to form a 3D dilated DS convolution. The key gains from replacing regular 3D convolutions with 3D DS convolutions can be seen in Table \ref{tab: depthwise}, where 3D DS convolutions reduce the number of parameters and multiply-accumulate operations (MACs) by a factor of $ \frac{K^3 \cdot N}{K^3 + N}$. For $K=3$ and $N=12$, a 3D DS convolution has $\sim8$ times less parameters and computations than a regular 3D convolution.\vspace{-2mm}

\begin{table}[!t]
  \caption{Efficiency: Regular 3D convolutions vs. 3D depthwise separable (DS) convolutions}
  \vspace{-1mm} 
  \label{tab: depthwise}
  \fontsize{7}{10}\selectfont
  \centering
  \begingroup
    \begin{tabular}{c|c|c}
    \hline
    \hline
    Layer & MACs  & Param. \\
    \hline
    3D Conv & $K^3 \cdot M \cdot H_o \cdot W_o \cdot D_o \cdot N$ & $K^3 \cdot M \cdot N$  \\
    3D DS Conv & $M \cdot H_o \cdot W_o \cdot D_o(K^3 + N)$ & $M(K^3 + N)$  \\
    \hline
    \hline
    \end{tabular}%
	\endgroup
	\vspace{-3mm}
\end{table}%

\subsection{Dilated residual dense block}

Incorporating multi-scale contextual information, including both local and global contexts, in FCNs is beneficial for accurate image segmentation \cite{Yu2015,Chen2017,Chen2018,Liu2015}. However, existing 3D FCN-based pulmonary lobe segmentation methods only make use of short-range local contexts due to their small-sized receptive fields. Since lung lobes are large and variable in size, and have a homogeneous appearance on CT, it is insufficient to segment them effectively based on such local information alone. Incomplete interlobar fissures and a wide spectrum of lung abnormalities further complicate the segmentation task. Long-range and global contexts are required to provide additional anatomical cues, such as the spatial configuration of surrounding structures and contextual relationships between them, to guide the lobar segmentation with the aim of reducing label ambiguity while enforcing spatial consistency. To capture this contextual information, we have to ensure our network has a large receptive field. Using large-sized convolutional kernels or adding more layers can enlarge the network receptive field but it increases memory and computational requirements. Another alternative is the use of striding or pooling but frequent downsampling can negatively impact the segmentation performance since detail information is decimated.\looseness=-1

\begin{figure}[!t]
    \setlength\abovecaptionskip{-0.1\baselineskip}
	\centering
	\includegraphics[scale=0.3]{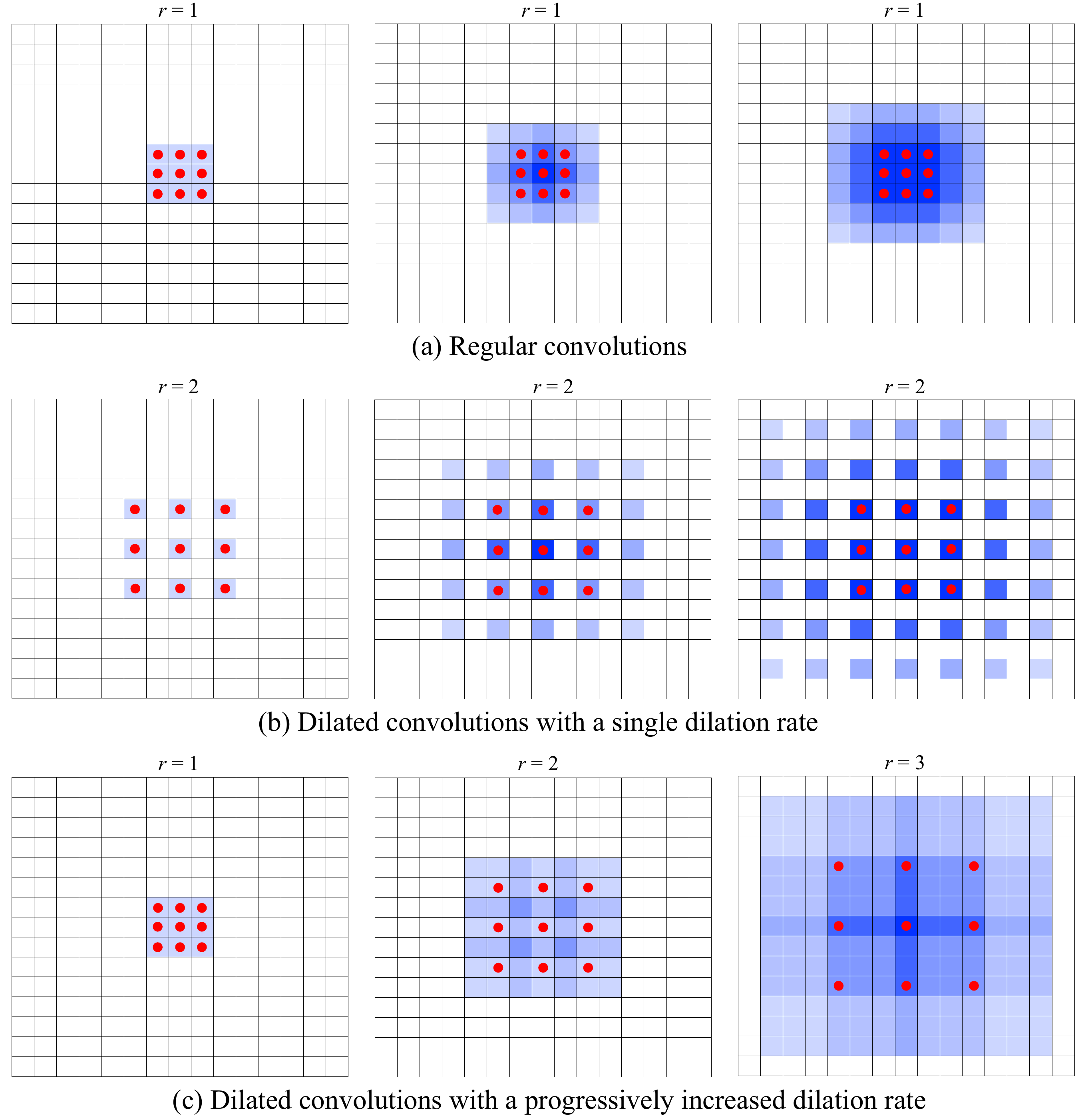}
	\caption{Receptive fields of regular convolutions and dilated convolutions in consecutive $3 \times 3 \times 3$ convolutional layers (from left to right). The blue voxels denote the receptive fields, with darker shades representing a higher frequency of overlaps between receptive fields of each kernel point (drawn in red). (a) Regular convolutions expand the receptive field linearly. (b) Dilated convolutions enlarge the receptive field exponentially but gridding artefacts appear when adopting a single dilation rate (e.g. $r=(2,2,2)$). (c) The problem can be alleviated using a progressively increased dilation rate (e.g. $r=(1,2,3)$) while maintaining a large receptive field at the top layer.}
	\label{fig: dilated}
	\vspace{-5mm} 
\end{figure}

To this end, we revisit the dilated convolutions presented by \citet{Chen2018}, which allow us to perform convolutions with dilated or upsampled kernels. A 3D dilated convolution with dilation rate $r\in\mathbb{Z}^{+}$ is a convolution in which the kernels are implicitly upsampled by inserting $r-1$ zeros between each kernel element, and it can be defined from (\ref{eq:1}) as follows:
\begin{equation}\label{eq:5}
\displaystyle Y_{h,w,d,n} = \sum\limits_{i,j,k,m} X_{h+ir,w+jr,d+kr,m} \cdot W_{i,j,k,m,n}
\end{equation}
When $r=1$, this is a regular 3D convolution. Figure \ref{fig: dilated}(c) shows $3 \times 3 \times 3$ convolutions with different dilation rates. A dilated convolution effectively expands the receptive field of kernels by adjusting the dilation rate, without introducing additional parameters and computations. However, the use of such sparse convolutional kernels ($r>1$) can cause gridding artefacts \cite{WangP2018} as the input is sampled in a checkerboard fashion. This could impair the consistency of local information and thus hamper the network performance. Moreover, when the dilation rates of convolutional layers arranging in cascade are the same or have a common factor relationship, such as $r=(2,2,2)$ or $(2,4,8)$, the gridding effect can be propagated to consecutive layers \cite{WangP2018} (Figure \ref{fig: dilated}(b)).\looseness=-1

\begin{figure}[!t]
	\setlength\abovecaptionskip{-0.2\baselineskip}
	\centering
	\includegraphics[scale=0.5]{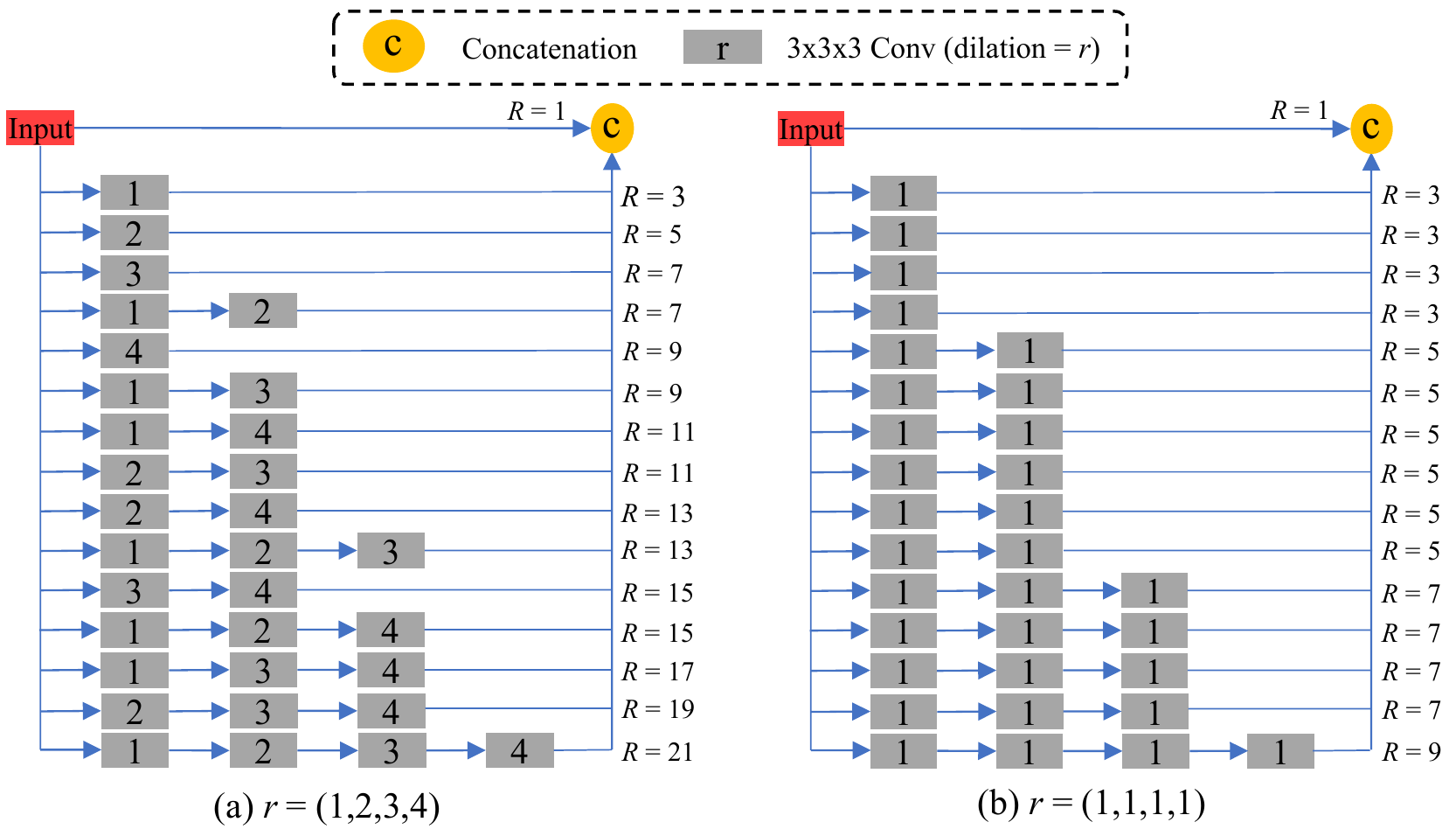}
	\caption{The unravelled view of dense connections. (a) Dense connectivity renders the cascade module of dilated convolutions to have a collection of $2^{4}$ paths for information propagation, each with a different subset of layers. This allows our DRDB to learn multi-scale context features with a large diversity of receptive fields, ranging from $R=1 \times 1 \times 1$ to $R=21 \times 21 \times 21$. (b) The typical dense block in DenseNet \cite{Huang2017} can also capture multi-scale context features but the scale diversity is limited and different layers may extract the same types of features multiple times, resulting in a certain redundancy \cite{ChenY2017}.}
	\label{fig: receptive field}
	\vspace{-5mm} 
\end{figure}

Previous works \cite{Yu2015, Chen2018, Chen2017} proposed modules that adopt dilated convolutions in cascade or in parallel to efficiently exploit contextual information for semantic segmentation. Suppose four $3 \times 3 \times 3$ convolutions with dilation rates $r=(1,2,3,4)$ are used, the cascade module exponentially enlarges the receptive field to capture large context. Nevertheless, the final output feature maps only contain single-scale contextual information encoded by a large, fixed receptive field of size $21^{3}$, which may not be optimal for segmenting objects of different sizes. In addition, classifying ambiguous voxels may require contextual information captured by diverse scales of receptive fields or a combination of them. On the other hand, the parallel module, which is built on the split-transform-merge principle, can capture image context at multiple scales but only with small-sized receptive fields (i.e. $3^{3}$, $5^{3}$, $7^{3}$ and $9^{3}$) and the scale diversity is limited by the number of layers. Motivated by the above observation, we devise a new module based on dilated convolutions \cite{Chen2018}, dense connections \cite{Huang2017} and residual learning \cite{He2016} to incorporate multi-scale contextual information for segmentation. We refer to this module as dilated residual dense block (DRDB), which is the basic building block of our PLS-Net. The DRDB not only inherits the advantages of both cascade and parallel modules of dilated convolutions, but it is also able to encode multi-scale context features of a large scale range in a dense manner, as seen in Figure \ref{fig: receptive field}(a).

\begin{figure}[!t]
	\setlength\abovecaptionskip{-0.1\baselineskip}
	\centering
	\includegraphics[scale=0.3]{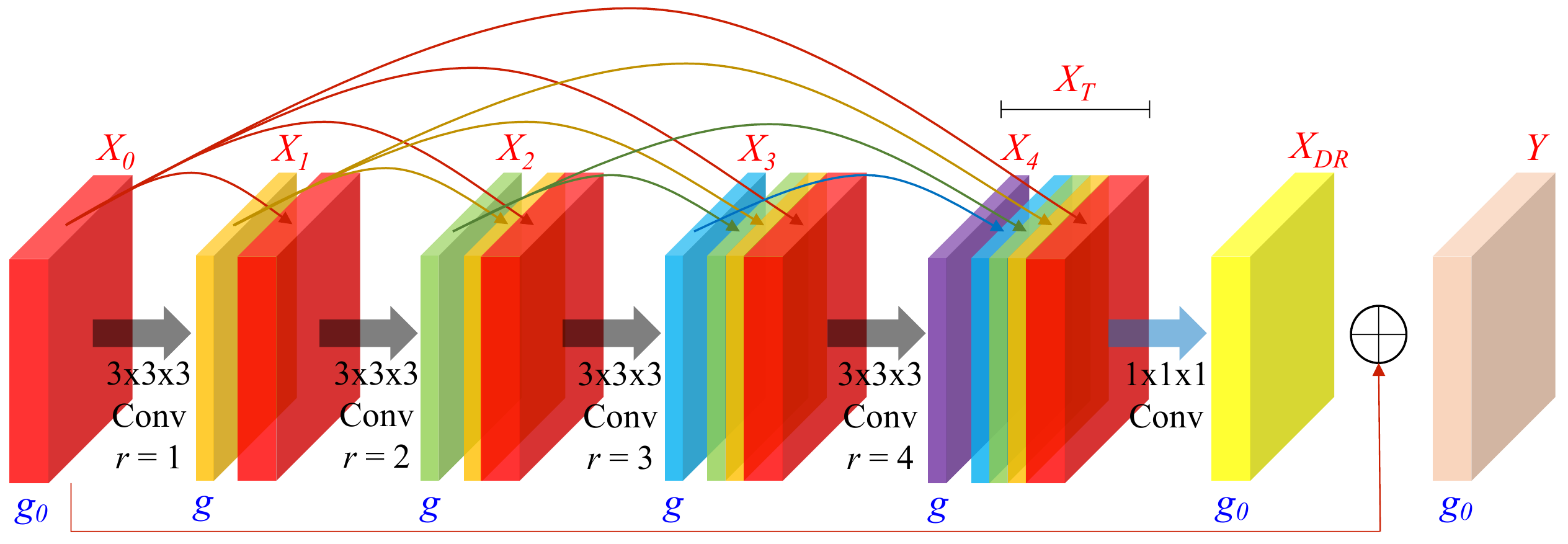}
	\caption{Dilated residual dense block (DRDB) architecture. We depict each block as a 4D feature map tensor; the thickness of each block indicates its relative number of channels. The hyperparameter $g$ is the growth rate, which regulates the amount of new features each dilated convolution layer adds to the cumulative information from preceding layers.}
	\label{fig: DRDB}
	\vspace{-5mm} 
\end{figure}

The structure of a DRDB is depicted in Figure \ref{fig: DRDB}. It consists of four densely connected $3 \times 3 \times 3$ convolution layers with dilation rates $r=(1,2,3,4)$, followed by a $1 \times 1 \times 1$ convolution layer and residual learning. A progressively increased dilation rate is used so that the receptive field of each dilated convolution layer arranged in cascade fully covers a cubic region without any holes, as shown in Figure \ref{fig: dilated}(c). In this way, the layers can work interdependently to alleviate the gridding problem caused by dilated convolutions while enlarging the receptive field exponentially at no extra parameters and computations. To incorporate multi-scale context features, we introduce dense connectivity \cite{Huang2017} to the cascade module of dilated convolutions, where each layer has direct connections to all subsequent layers. Such connections strengthen feature propagation and promote feature reuse, as early-layer features can be accessed by all other layers via recursive concatenations. Thus, the $i$-th dilated convolution layer, $i=\{1,2,3,4\}$, of the DRDB receives the feature maps of all preceding layers as input, and its output can be computed as:
\begin{equation}\label{eq:6}
\displaystyle X_{i} = H_{3,r_{i}}([X_0,X_1,...,X_{i-1}])
\end{equation}
where $H_{3,r_{i}}(\cdot)$ denotes the function of a $3 \times 3 \times 3$ convolution with dilation rate $r_{i}$ (the dilation rate of the layer $i$), and $[X_0,X_1,...,X_{i-1}]$ represents the concatenation of the input feature maps $X_0$ and feature maps produced by the dilated convolution layers $1,2,...,i-1$. Through a series of dilated convolutions and feature concatenations, the output $X_T$ of the DRDB is the collective concatenation of each layer's feature maps:
\begin{equation}\label{eq:7}
\displaystyle X_{T} = [X_0,X_1,...,X_{4}]
\end{equation}
If $X_0$ has $g_0$ feature maps and each dilated convolution layer generates $g$ feature maps (where $g$ is the growth rate \cite{Huang2017} and is typically set to a small value between 8 and 48), it will have $g_0+4g$ feature maps encoded with multi-scale contextual information. Figure \ref{fig: receptive field}(a) shows the unraveled view of dense connectivity in a DRDB to illustrate the scale diversity of the extracted features. Intuitively, the obtained output can be seen as a sampling of the input with different scale of receptive fields or convolutional kernels. Compared to the basic dense block in \cite{Huang2017} (Figure \ref{fig: receptive field}(b)), our DRDB that adopts dilated convolutions in combination with dense connections is able to capture features in a larger range of scales whilst mitigating the feature redundancy issue introduced by dense connectivity \cite{ChenY2017}.\looseness=-1

To improve the computational and parameter efficiency, a $1 \times 1 \times 1$ convolution is then applied to project the high-dimensional feature maps $X_{T}$ onto a low-dimensional space:
\begin{equation}\label{eq:8}
\displaystyle X_{DR} = H_{1}(X_T)
\end{equation}
This reduces the channel dimension of $X_{T}$ from $g_0+4g$ to $g_0$ and enables the introduction of residual learning \cite{He2016}, via a shortcut connection and element-wise addition between the input and output of the DRDB, to further enhance the information flow and refine the output feature representations. The final output of the DRDB can be obtained by:
\begin{equation}\label{eq:9}
\displaystyle Y = X_{DR}+X_0
\end{equation}

A naive implementation of the DRDB can impose significant GPU memory burden during training; convolutional and intermediate feature maps (outputs of the concatenation, addition or batch normalisation (BN) operations) generated in each layer are kept in memory for back-propagation to compute gradients, resulting in a quadratic memory growth with network depth. Since these intermediate outputs responsible for most memory usage are cheap to compute, a memory-efficient DRDB can be implemented by storing only the convolutional outputs during the forward pass. The intermediate outputs are discarded after use in the forward pass and recomputed on-the-fly as necessary during back-propagation. This memory-efficient DRDB implementation reduces the feature map memory consumption from quadratic to linear; the memory savings allow us to train our PLS-Net, which is mainly built from the DRDBs, on a reasonable memory budget, with only a small increase (around $15\%$) in training time. The idea of recomputing intermediate feature maps for saving memory has also been explored in \cite{Pleiss2017}.\vspace{-3mm}

\subsection{Network architecture}

\begin{figure*}[!t]
	\setlength\abovecaptionskip{-0.1\baselineskip}
	\centering
	\includegraphics[scale=0.55]{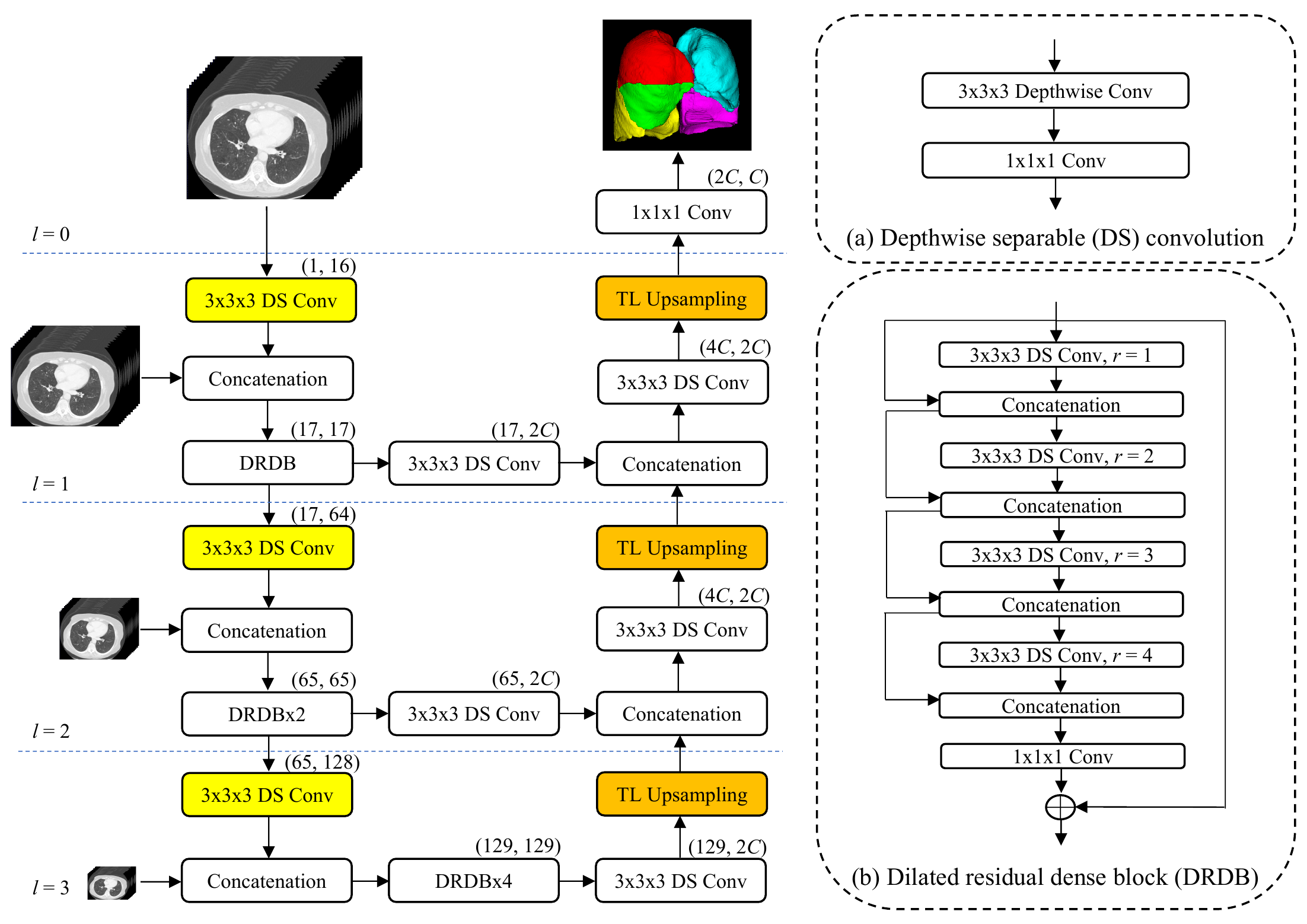}
	\caption{The architecture of our proposed pulmonary lobe segmentation network (PLS-Net). It is based on an asymmetrical encoder-decoder structure with (a) depthwise separable (DS) convolution layers, (b) dilated residual dense blocks (DRDBs) with a growth rate of $g=12$, and input reinforcement at each downsampled resolution. The number of input and output channels is denoted on the top of each convolutional layer. Yellow and orange boxes represent layers that perform downsampling (with strided DS convolution) and upsampling (with trilinear (TL) interpolation) operations, respectively. \textit{l} indicates the resolution level, \textit{r} represents the dilation rate, and \textit{C} denotes the number of output classes.}
	\label{fig: architecture}
	\vspace{-3mm} 
\end{figure*}

As illustrated in Figure \ref{fig: architecture}, we design our PLS-Net based on three insights, i.e. efficiency, multi-scale feature representations, and high-resolution 3D input/output. The PLS-Net is an asymmetric encoder-decoder based network, consisting of (i) a deep encoder that extracts multi-scale context features from a high-resolution volumetric CT input image, and (ii) a shallow, light-weight decoder that decodes these features to generate correspondingly-sized pulmonary lobar segmentations.

The encoder consists of stacked DRDBs, with three convolutional layers with stride 2 to increase the number of feature maps. The growth rate of each DRDB is empirically set to $g=12$, which gives a good trade-off between accuracy, memory footprint and computational efficiency (see Section \ref{sec:ablation-study}). According to the \textit{unravelled view} of skip connections presented by \citet{Veit2016}, the encoder can intuitively be viewed as an ensemble of variable depth and receptive field networks. The ensemble of different information propagation paths throughout the network can capture and aggregate multi-scale anatomical contexts in the CT images. To mitigate the spatial information loss due to convolutional and downsampling operations, we introduce an input reinforcement (IR) scheme, where the input image is reintroduced after each downsampling layer of the encoder network. Specifically, the input image is downsampled by a factor of $2^l$ via trilinear interpolation and concatenated with the output of the strided convolutional layer in the resolution level $l$, where $l=\{1,2,3\}$. The IR scheme helps in retaining spatial information throughout the encoding process.\looseness=-1

The spatial resolution of the output feature maps produced by the encoder is $\frac{1}{8}$ of the input image resolution. A simple decoder that directly upsamples the feature maps back to the input resolution would generate a coarse output with reduced boundary adherence \cite{Shelhamer2017}. Inspired by \cite{Imran2018,Ronneberger2016}, we construct the decoder based upon a \textit{conv-upsample-merge} principle: (i) \textit{conv}: the feature maps from  resolution levels $l$ and $l-1$ are convolved with $3 \times 3 \times 3$ voxel kernels to produce $2C$ feature maps, where $C$ represents the number of output classes (in this study $C=6$, i.e. five lung lobes and background); (ii) \textit{upsample}: the $2C$ feature maps from resolution level $l$ are upsampled by a factor of 2 using trilinear interpolation, which is fast to compute and avoids checkerboard artefacts induced by transposed convolutions \cite{odena2016}; and (iii) \textit{merge}: the upsampled feature maps are then concatenated with the $2C$ feature maps from resolution level $l-1$. This process is repeated until the spatial resolution of the feature maps is the same as the input image. Finally, the feature maps are fed to a $1 \times 1 \times 1$ convolutional layer and a softmax activation function to yield $C$ probability maps corresponding to each of the classes. The final predicted segmentation corresponds to the class with maximum probability at each voxel.

The convolutional kernel sizes used in the PLS-Net are either $1 \times 1 \times 1$ or $3 \times 3 \times 3$ voxels. To further reduce the computational cost and number of parameters, we adopt $3 \times 3 \times 3$ DS convolutions in place of regular $3 \times 3 \times 3$ convolutions; each of which consists of a $3 \times 3 \times 3$ depthwise convolution and a $1 \times 1 \times 1$ convolution. All convolutional layers in the network are followed by BN and a ReLU activation function, except for the depthwise convolutional layers and the final layer. Convolutional layers in each resolution level $l$ use zero-padding to maintain the feature map size.

\subsection{Implementation details}

\subsubsection{Pre- and post-processing}
To handle CT images with varying voxel sizes, we first standardised them by resampling to an isotropic voxel size of $1$ mm$^{3}$. These resampled images, with dimensions ranging from $352 \times 352 \times 344$ to $400 \times 400 \times 396$ voxels, were then normalised to zero mean and unit variance, and used as training or testing input of the PLS-Net. The fully convolutional nature of our PLS-Net allows it to process CT input volumes of different sizes and produce segmentation maps of corresponding dimensions. Since training and inference were performed in the resampled resolution, the segmentation outputs were resampled back to their original image resolution for performance evaluation.

\subsubsection{Training} 
We implemented our PLS-Net using the Pytorch library \cite{paszke2017automatic}. To reduce the memory footprint for processing high-resolution CT input volumes, we trained the PLS-Net using the mixed-precision training strategy \cite{Paulius2018}, which provided memory savings and improved throughput of half-precision (16-bit floating point) training whilst maintaining the accuracy of the conventional single-precision (32-bit floating point) training. The network was trained from scratch with weights initialised from a Gaussian distribution ($\mu=0$, $\sigma=0.01$). The batch size was set to 1 to enable training with different input sizes. We used the categorical cross-entropy as the network loss function, and the Adam optimiser (learning rate $= 0.001$, $\beta_1=0.9$, $\beta_2=0.999$). Each training session required about 8.5 GB of GPU memory, and the training was stopped when there was no improvement in the model loss on the validation set over 20 epochs.\vspace{-3mm} \looseness=-1

\section{Experiments \& Results}

Two sets of experiments were conducted on the multi-institutional dataset described in Section \ref{sec:materials} to validate the effectiveness and efficiency of our proposed PLS-Net. The first set aimed to compare the performance of the PLS-Net against five state-of-the-art methods, including four FCN-based methods (P-HNN+RW \cite{George2017}, FRV-Net \cite{Ferreira2018}, 3D U-Net \cite{Park2019} and PDV-Net \cite{Imran2018}), and a classical method (PTK \cite{Doel2012}) based on a fissureness filter combining image appearance and anatomical information from pulmonary airways and vessels. Note that PDV-Net \cite{Imran2018} has been shown to be comparable or better than the methods proposed in \cite{Lassen2013,Bragman2017,Giuliani2018}. The compared methods were implemented either by ourselves following the papers or using the source code provided by the authors. All methods were evaluated using ten-fold cross-validation except for PTK, which is an unsupervised approach. Specifically, the dataset was split into ten fixed folds, each of which consisted of 10, 8 and 3 CT images from the LTRC, LIDC-IDRI and COPD databases, respectively. Each fold was then iteratively used once for testing while the remaining nine folds were used for training and validation (with a ratio of 8:1). We measured the segmentation accuracy by computing the Dice similarity coefficient (DSC) and average symmetric surface distance (ASD), and reported the mean and standard deviation over all 210 cases. The second set of experiments (ablation experiments) aimed at analysing the performance and efficiency gains contributed by each proposed component of the PLS-Net. Different PLS-Net variants were trained and evaluated following the same ten-fold cross-validation protocol. All experiments were carried out on a machine with an Intel Xeon E5-2600 CPU and an NVIDIA GTX 1080 Ti GPU.\vspace{-3mm} \looseness=-1

	\subsection{Evaluation metrics}
	
	The DSC and ASD between automatic and expert-delineated reference segmentations were used for quantitative evaluation of segmentation performance. The DSC measures the relative volume overlap between two segmentations and is defined as:
	\begin{equation}\label{eq:10}
	\displaystyle DSC(A,B) = \frac{2 \vert A \cap B \vert}
	{\vert A \vert + \vert B \vert}
	\end{equation}
	where $A$ and $B$ denote the automatic and reference segmentations, respectively. The ASD is a symmetric measure of average minimum distance between surface voxels of two segmentations and is expressed as:	
	\begin{equation}\label{eq:11}
	\displaystyle ASD(A,B) =  \frac{\sum\limits_{a\in S_A} \min\limits_{b\in S_B}d(a, b) + \sum\limits_{b \in S_B} \min\limits_{a\in S_A}d(b, a)}{\vert S_A\vert + \vert S_B\vert}
	\end{equation}
	with $d(\cdot)$ being the Euclidean distance while $S_A$ and $S_B$ are the surface voxels of automatic and reference segmentations, respectively. Each metric highlights different aspects of segmentation quality, allowing a more comprehensive assessment of a method. A Wilcoxon signed-rank test with a significance level of 0.05 was employed to assess whether the performance difference between two segmentation methods was statistically significant. For FCN-based approaches, we also reported the number of parameters and MACs (per forward pass) to show the model size and its computational complexity.\vspace{-3mm}

	\subsection{Comparison with the state-of-the-art}
	
		\begin{table*}[!t]	
		\caption{Quantitative comparison with state-of-the-art methods in terms of MACs, parameters, DSC, ASD, average training and testing time. DSC and ASD metrics are given in the form of (mean $ \pm $ standard deviation), and ASD is in millimetres. Boldface denotes the best metric value, and an asterisk indicates a significant difference (\textit{p} $<$ 0.05) compared with other methods}
		\vspace{-1mm} 
		\label{tab: result_quantitative}
		\fontsize{6}{7.5}\selectfont
		\centering
		\begin{tabular}{c|c|c|c|c|c|c|c|c|c|c|c}
			\hline
			\hline
			Method & MACs & Param. & Metric & RUL   & RML   & RLL   & LUL    & LLL & Overall  & Train & Test \\
			\hline
			\multirow{2}[1]{*}{PTK \cite{Doel2012}}& \multirow{2}[1]{*}{-} & \multirow{2}[1]{*}{-} & DSC  & 0.872 $ \pm $ 0.141 & 0.792 $ \pm $ 0.175 & 0.861 $ \pm $ 0.148 & 0.870 $ \pm $ 0.124 & 0.857 $\pm$ 0.130 & 0.850 $ \pm $ 0.148 & \multirow{2}[1]{*}{-} & \multirow{2}[1]{*}{613 s}\\
			&       &       & ASD    & 2.479 $ \pm $ 1.232 & 3.175 $ \pm $ 1.287 & 2.829 $ \pm $ 1.394 & 2.588 $ \pm $ 1.284 & 2.629 $ \pm $ 1.216 & 2.740 $ \pm $ 1.305  & & \\
			\hline
			\multirow{2}[0]{*}{P-HNN+RW \cite{George2017}} & \multirow{2}[0]{*}{67.43B} & \multirow{2}[0]{*}{14.72M} & DSC  & 0.904 $ \pm $ 0.070 & 0.840 $ \pm $ 0.135 & 0.876 $ \pm $ 0.117 & 0.901 $ \pm $ 0.080 & 0.910 $ \pm $ 0.065 & 0.886 $ \pm $ 0.101 & \multirow{2}[1]{*}{10 h} & \multirow{2}[1]{*}{314 s}\\
			&       &       & ASD    & 1.333 $ \pm $ 0.990 & 1.878 $ \pm $ 1.319 & 1.646 $ \pm $ 1.127 & 1.219 $ \pm $ 0.914 & 1.226 $ \pm $ 0.875 & 1.460 $ \pm $ 1.086 &  & \\
			\hline
			\multirow{2}[0]{*}{FRV-Net \cite{Ferreira2018}} & \multirow{2}[0]{*}{161.34B} & \multirow{2}[0]{*}{19.11M} & DSC  & 0.938 $ \pm $ 0.053 & 0.899 $ \pm $ 0.079 & 0.940 $ \pm $ 0.044 & 0.946 $ \pm $ 0.037 & 0.932 $ \pm $ 0.067 & 0.933 $ \pm $ 0.060 & \multirow{2}[1]{*}{64 h} & \multirow{2}[1]{*}{48.7 s} \\
			&       &       & ASD    & 0.914 $ \pm $ 0.618 & 1.441 $ \pm $ 1.143 & 0.946 $ \pm $ 0.463 & 0.843 $ \pm $ 0.480 & 0.875 $ \pm $ 0.549 & 1.004 $ \pm $ 0.730 & & \\
			\hline
			\multirow{2}[0]{*}{3D U-Net \cite{Park2019}} & \multirow{2}[0]{*}{645.66B} & \multirow{2}[0]{*}{18.08M} & DSC  & 0.947 $ \pm $ 0.046 & 0.895 $ \pm $ 0.117 & 0.949 $ \pm $ 0.044 & 0.960 $ \pm $ 0.029 & 0.946 $ \pm $ 0.046 & 0.940 $ \pm $ 0.067 & \multirow{2}[1]{*}{71 h} & \multirow{2}[1]{*}{151 s}\\
			&       &       & ASD    & 0.940 $ \pm $ 0.666 & 1.561 $ \pm $ 1.340 & 0.982 $ \pm $ 0.667 & 0.843 $ \pm $ 0.426 & 0.869 $ \pm $ 0.569 & 1.039 $ \pm $ 0.840 & & \\
			\hline
			\multirow{2}[0]{*}{PDV-Net \cite{Imran2018}} & \multirow{2}[0]{*}{292.13B} & \multirow{2}[0]{*}{0.90M} & DSC  & 0.951 $ \pm $ 0.049 & 0.904 $ \pm $ 0.070 & 0.949 $ \pm $ 0.048 & 0.957 $ \pm $ 0.033 & 0.944 $ \pm $ 0.038 & 0.940 $ \pm $ 0.054 & \multirow{2}[1]{*}{35 h} & \multirow{2}[1]{*}{2.5 s}\\
			&       &       & ASD    & 0.830 $ \pm $ 0.582 & 1.389 $ \pm $ 1.139 & 0.818 $ \pm $ 0.526 & 0.786 $ \pm $ 0.473 & 0.839 $ \pm $ 0.578 & 0.932 $ \pm $ 0.738 &  & \\
			\hline
			\multirow{2}[1]{*}{PLS-Net} & \multirow{2}[1]{*}{103.69B} & \multirow{2}[1]{*}{0.25M} & DSC  & \textbf{0.962 $ \pm $ 0.045*} & \textbf{0.936 $ \pm $ 0.059*} & \textbf{0.963 $ \pm $ 0.043*} & \textbf{0.968 $ \pm $ 0.028*} & \textbf{0.961 $ \pm $ 0.036*} & \textbf{0.958 $ \pm $ 0.045*} & \multirow{2}[1]{*}{23 h} & \multirow{2}[1]{*}{0.8 s}\\
			&       &       & ASD    & \textbf{0.704 $ \pm $ 0.557*} & \textbf{1.143 $ \pm $ 0.969*} & \textbf{0.673 $ \pm $ 0.433*} & \textbf{0.650 $ \pm $ 0.408*} & \textbf{0.715 $ \pm $ 0.541*} & \textbf{0.777 $ \pm $ 0.641*} & & \\
			\hline
			\hline
			\multicolumn{9}{@{}l}{RUL: Right upper lobe, RML: Right middle lobe, RLL: Right lower lobe, LUL: Left upper lobe, LLL: Left lower lobe.}
		\end{tabular}%
		\vspace{-3mm} 
	\end{table*}%

\begin{figure}[!t]
		\setlength\abovecaptionskip{-0.1\baselineskip}
		\centering
		\includegraphics[width=\linewidth]{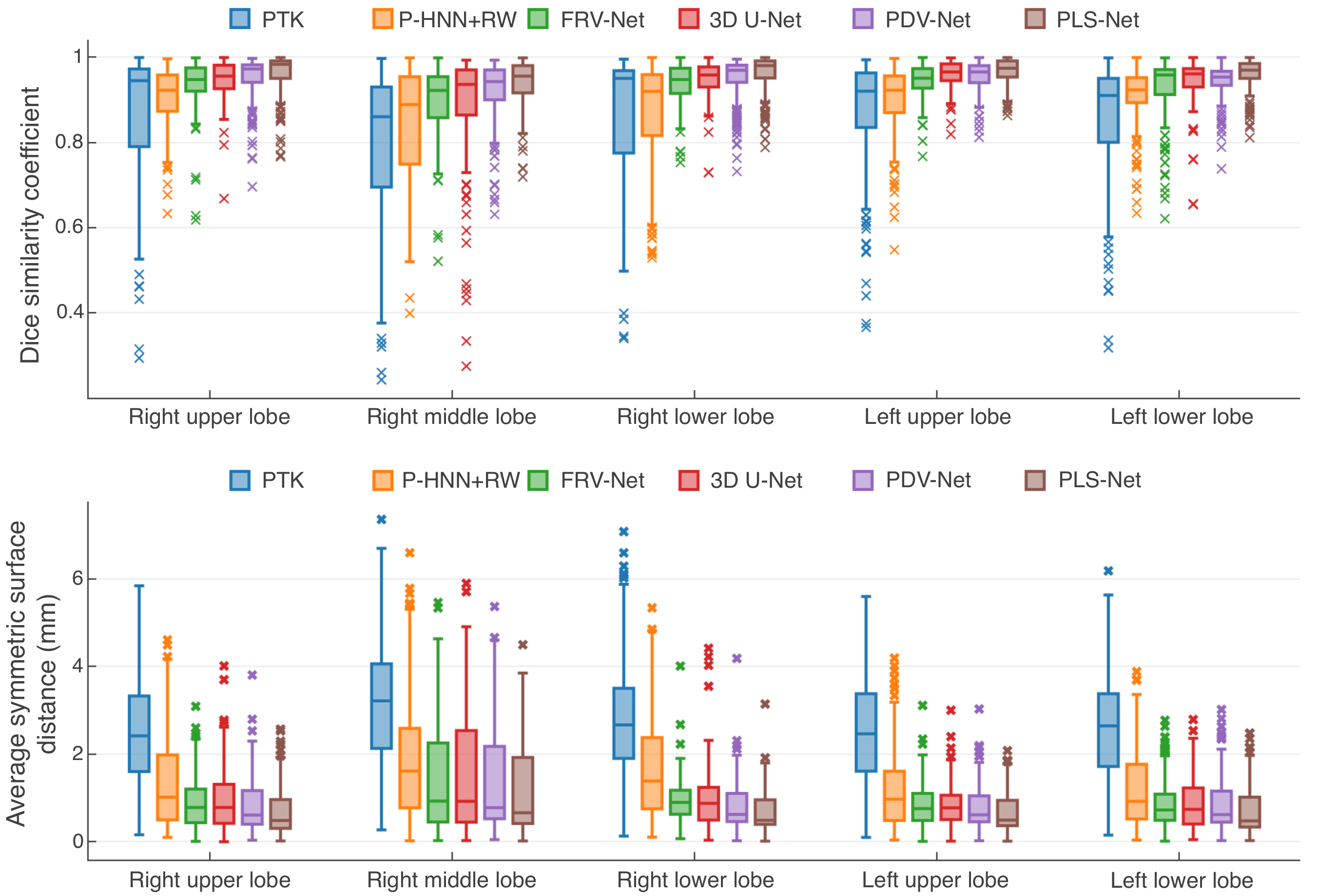}
		\caption{Box and whisker plots of Dice similarity coefficient and average symmetric surface distance over all lobar structures for each method.}
		\label{fig: result_boxplot}
		\vspace{-5mm}
	\end{figure}

	Table \ref{tab: result_quantitative} reports the quantitative results of experiments comparing our proposed PLS-Net with state-of-the-art pulmonary lobe segmentation methods. The box and whisker plots of DSC and ASD are shown in Figure \ref{fig: result_boxplot}. From the results, it can be seen that our PLS-Net significantly outperformed the compared methods in segmenting each of the lung lobes ($p<0.05$), yielding a pooled mean DSC of $0.958\pm0.045$ and ASD of $0.777\pm0.641$ mm. The standard deviations of the metrics produced by our method were also lower than those of other approaches, indicating the robustness and generalisation capability of our method. Moreover, the PLS-Net demonstrated better efficiency over the competing methods; it required only 0.25 million parameters and an average of 104 billion MACs to achieve state-of-the-art performance, with an average runtime of 0.8 and 9.7 seconds per CT image on the GPU and CPU, respectively. The training time of the PLS-Net was also faster than most of the FCN-based methods except P-HNN+RW, whose network was fine-tuned from the VGG-16 model pre-trained on the ImageNet dataset. There was no statistically significant difference between the overall performance of the other 3D FCN-based methods (FRV-Net, 3D U-Net and PDV-Net) ($p>0.05$); however, the PDV-Net, which also uses densely-connected narrow layers and processed CT scans in a whole-volume fashion (but at a downsampled resolution) instead of patch-wise, required fewer parameters to obtain the comparable performance and was faster in both training and testing time.\looseness=-1
	
 	In addition, the performance of the P-HNN+RW that combines a 2D FCN and a 3D random walker was significantly below those of 3D FCN-based methods ($p<0.05$), corroborating the efficacy of 3D FCNs in segmenting the lung lobes. Furthermore, all the FCN-based methods performed significantly faster and better than the classical unsupervised PTK method employing hand-crafted features and prior segmentations of lung structures ($p<0.05$). This suggests that FCNs can learn more discriminative features in a data-driven manner to better handle the high variability of pulmonary lobar anatomy in patients with lung diseases. It is worth noting that the segmentation accuracy for the right middle lobe (RML) was notably lower than the other lobes across all the comparison methods ($p<0.05$), revealing that the RML segmentation is relatively more challenging. This may be owing to its relatively small size and large anatomical variations; the prevalence of incomplete right major and minor fissures further complicates the task. Nevertheless, the largest performance improvement was observed for the RML when comparing our method with the top-performing approaches. This finding demonstrates that our PLS-Net with the deeper high-resolution multi-scale 3D features is more effective for the RML segmentation problem.
 	
\begin{figure*}[!t]
	\setlength\abovecaptionskip{-0.3\baselineskip}
	\centering
	\includegraphics[scale=0.5]{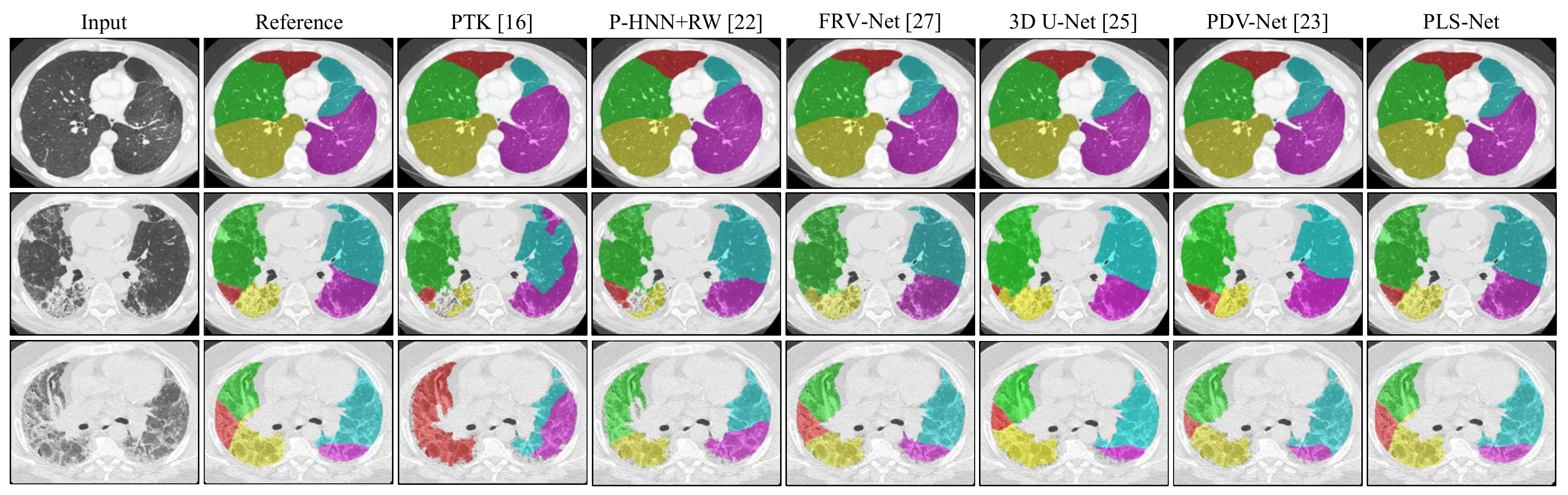}
	\caption{Qualitative comparison with state-of-the-art methods. Each row represents an individual case, with mild, moderate and severe pathologies listed from top to bottom. The columns from left to right are the axial slices of the input CT volumes, expert-based reference segmentations, segmentations of the PTK \cite{Doel2012}, P-HNN+RW \cite{George2017}, FRV-Net \cite{Ferreira2018}, 3D U-Net \cite{Park2019}, PDV-Net \cite{Imran2018}, and our proposed PLS-Net. The right upper lobe, right middle lobe, right lower lobe, left upper lobe and left lower lobe are depicted in red, green, yellow, cyan and magenta, respectively.}
	\label{fig: result_qualitative}
	\vspace{-5mm} 
	\end{figure*}

 	Figure \ref{fig: result_qualitative} shows the segmentation results of different methods on representative cases with various levels of pathology. As can be seen, the lobe segmentation is challenging given the structural variability across patients, either naturally or caused by diseases. The lobar boundaries are hard to identify due to the incompleteness or fuzzy appearance of fissures, and they can be confused by pathologies (middle and bottom row) or other fissure-resembling structures (e.g. accessory fissures (top row)). Nonetheless, our PLS-Net can still accurately delineate the lung lobes, demonstrating better effectiveness and robustness over the other methods. \vspace{-2mm}

	\subsection{Ablation studies}\label{sec:ablation-study}
	
	We investigated the effects of the three key components of our PLS-Net: depthwise separable (DS) convolutions, dilated residual dense blocks (DRDBs) and input reinforcement (IR). Quantitative results of the ablation study are summarised in Table \ref{tab: ablation_study_PLS_Net}. The baseline network PLS\_DS0DRDB0IR0 was built without employing DS convolutions, DRDBs or IR: standard $3 \times 3 \times 3$ convolutions were used, and DRDBs were replaced by regular convolutional layer stacks (without shortcut connections and dilated convolutions, i.e. $r=(1,1,1,1)$). It performed poorly with a pooled mean DSC of $0.813\pm0.094$ and ASD of $1.478\pm1.116$ mm. The poor performance may be due to the difficulty of training caused by the vanishing gradient problem \cite{He2016}, leading the network to prematurely converge to an inferior solution. This suggests that simply stacking many narrow layers (e.g. 12 kernels per layer) to form a very deep 3D FCN with a high-resolution input image would not result in good performance. Replacing regular convolutional layer stacks with DRDBs and/or introducing the IR scheme yielded a statistically significant performance improvement over the baseline ($p<0.05$), with the PLS\_DS0DRDB1IR1 that contains both components achieving the best performance. For all network architectures, adopting $3 \times 3 \times 3$ DS convolutions in place of $3 \times 3 \times 3$ convolutions reduced the network parameters and computational complexity (in terms of MACs) by 83\% to 91\% at the expense of only a small reduction in accuracy. Our PLS-Net with DS convolutions was nearly as accurate as the PLS\_DS0DRDB1IR1 built with regular convolutions ($p>0.05$) while requiring 6$\times$ fewer parameters and being computationally cheaper to train and run. In practice, the average training/inference time on the GPU was reduced from 44 h/4.2 s to 23 h/0.8 s when using DS convolutions. These analyses demonstrate the effectiveness and benefits of each proposed component of the PLS-Net on its own or in combination with others.\looseness=-1
	
	\begin{table}[!t]
		\fontsize{6}{7.5}\selectfont
		\caption{Ablation study of the proposed PLS-Net. DS: adopting depthwise separable convolutions. DRDB: employing dilated residual dense blocks. IR: with input reinforcement at each downsampled resolution}
		\vspace{-1.5mm}
		\label{tab: ablation_study_PLS_Net}
		\centering
		\begingroup
		\setlength{\tabcolsep}{5pt} 
		\renewcommand{\arraystretch}{1}
		\begin{tabular}{c|c|c|c|c|c|c}
			\hline
			\hline
			DS    & DRDB    & IR    & MACs  & Param. & mDSC  & mASD (mm) \\
			\hline
			&       			&      			& 401.34B & 0.65M & 0.813 $ \pm $ 0.094 & 1.478 $ \pm $ 1.116 	\\
			\checkmark      &       			&       		& 55.48B  & 0.06M & 0.805 $ \pm $ 0.093 & 1.495 $ \pm $ 1.110 	\\
			& \checkmark     	&       		& 651.87B & 1.45M & 0.947 $ \pm $ 0.046{*} & 0.823 $ \pm $ 0.742{*}   \\
			&       			& \checkmark    & 407.64B & 0.65M & 0.842 $ \pm $ 0.086{*} & 1.272 $ \pm $ 0.982{*} 	\\
			\checkmark      & \checkmark     	&       		& 101.31B  & 0.25M & 0.944 $ \pm $ 0.049{*} & 0.832 $ \pm $ 0.749{*} 	\\
			\checkmark      &       			& \checkmark    & 56.21B  & 0.06M & 0.833 $ \pm $ 0.088{*} & 1.295 $ \pm $ 0.988   \\
			& \checkmark   		& \checkmark    & 666.36B & 1.47M & 0.960 $ \pm $ 0.046{*} & 0.770 $ \pm $ 0.643{*}   \\
			\checkmark      & \checkmark     	& \checkmark    & 103.69B  & 0.25M & 0.958 $ \pm $ 0.045{*} & 0.777 $ \pm $ 0.641{*}   \\
			\hline
			\hline
			\multicolumn{7}{@{}l}{* indicates a significant difference $(p < 0.05)$ compared with the baseline.}
		\end{tabular}%
		\endgroup
		\vspace{-5mm} 
	\end{table}%
	
	It can be observed that the use of DRDBs contributes the most to the performance gains of the PLS-Net. We attribute the increase in accuracy to the effective design of the DRDB that combines the advantages of dilated convolutions, dense connectivity and residual learning. To verify the contribution of each of these elements, we conducted another ablation analysis; we set the PLS\_DS1DRDB0IR1 ($6^{th}$ combination in Table \ref{tab: ablation_study_PLS_Net}) as the baseline network and then added the elements individually or collectively to its convolutional layer stacks. As can be seen in Table \ref{tab: ablation_study_DRDB}, each element significantly improved the performance of the baseline ($p<0.05$), as the resulting networks can benefit from the enlarged receptive field offered by dilated convolutions or the improved information and gradient flow induced by dense or residual connections. Although the top layer of the baseline network already have large theoretical receptive fields almost covering the entire input image, the effective size of such fields could be much smaller in practice \cite{Luo2016,Liu2015}. This may be the reason why improving the use of context with dilated convolutions can yield better performance and also suggests that global information is necessary for accurate pulmonary lobe segmentation. The deploying of feature extraction blocks with two elements would generally perform better than with only one. Our PLS-Net built with DRDBs gave the best performance, justifying our design choice of the DRDB.\looseness=-1
	
	\begin{table}[!t]
		\fontsize{6}{7.5}\selectfont
		\caption{Ablation investigation of each component of the dilated residual dense block. DL: with dilated convolutions. DC: with dense connections. RL: with residual learning}
		\vspace{-1.5mm}
		\label{tab: ablation_study_DRDB}
		\centering		
		\begin{tabular}{c|c|c|c|c|c|c}
			\hline
			\hline
			DL    & DC    & RL    & MACs  & Param. & mDSC  & mASD (mm) \\
			\hline
			&       			&      			& 56.21B & 0.06M & 0.833 $ \pm $ 0.088 & 1.295 $ \pm $ 0.988 \\
			\checkmark     &       			&       		& 56.21B & 0.06M & 0.863 $ \pm $ 0.078{*} & 1.072 $ \pm $ 0.816{*} \\
			& \checkmark     &       		& 103.69B & 0.25M & 0.901 $ \pm $ 0.058{*} & 0.887 $ \pm $ 0.853{*} \\
			&       			& \checkmark    & 56.21B & 0.06M & 0.856 $ \pm $ 0.087{*} & 1.076 $ \pm $ 0.916{*} \\
			\checkmark     & \checkmark     &       		& 103.69B & 0.25M & 0.938 $ \pm $ 0.051{*} & 0.792 $ \pm $ 0.679{*} \\
			\checkmark     &       			& \checkmark    & 56.21B & 0.06M & 0.902 $ \pm $ 0.077{*} & 0.946 $ \pm $ 0.812{*} \\
			& \checkmark   	& \checkmark    & 103.69B & 0.25M & 0.936 $ \pm $ 0.048{*} & 0.812 $ \pm $ 0.701{*} \\
			\checkmark     & \checkmark     & \checkmark    & 103.69B & 0.25M & 0.958 $ \pm $ 0.045{*} & 0.777 $ \pm $ 0.641{*} \\
			\hline
			\hline
			\multicolumn{7}{@{}l}{* indicates a significant difference $(p < 0.05)$ compared with the baseline.}
		\end{tabular}%
		\vspace{-5mm} 
	\end{table}%
		
	The proposed PLS-Net contains 7 DRDBs, each with a growth rate of $g=12$ and a progressively increased dilation rate of $r=(1,2,3,4)$; such modular architecture allows flexible management of the network's parameters, computational cost, and memory footprint by adjusting the growth rate to meet the hardware constraints and task complexity. We analysed the influence of changing these hyperparameters on the PLS-Net performance. From Table \ref{tab: ablation_study_hyperparameters}, we can see that increasing the growth rate $g$ to 18 or 24 yielded only marginal gains in performance but increased computational and memory costs. Whilst reducing the growth rate to $g=8$ yielded a smaller and more efficient model, it also led to a proportional loss in accuracy. As we used a different dilation rate pattern (e.g. $r=(2,2,2,2)$ or $r=(1,2,4,8)$ that has been commonly used in previous works \cite{Yu2015,Chen2017,Chen2018}) for the DRDB with $g=12$, the segmentation accuracy was clearly lower than that of $r=(1,2,3,4)$. This could be due to the gridding issue caused by dilated convolutions and also underlines the effectiveness of using a progressively increased dilation rate. Moreover, we found that deeper variants of the PLS-Net (with additional DRDBs in the resolution level $l=3$) did not show better performance. One possible reason is that the receptive field of the PLS-Net is already sufficient for the task.\looseness=-1\vspace{-2mm}

	\begin{table}[!t]
	  \fontsize{6}{7.5}\selectfont
	  \caption{Effect of hyperparameters: Dilation rate $r$ \& Growth rate $g$}
	  \vspace{-1.5mm} 
	  \label{tab: ablation_study_hyperparameters}
	  \centering
	    \begin{tabular}{c|c|c|c|c|c|c}
	    \hline
	    \hline
	    $r$     & $g$     & MACs  & Param. & Memory & mDSC  & mASD (mm) \\
	    \hline
	    1-2-3-4 & 8     & 87.22B & 0.22M & 7.51 GB & 0.947 $ \pm $ 0.054{*} & 0.831 $ \pm $ 0.715 \\
	    1-2-3-4 & 12    & 103.69B & 0.25M & 8.39 GB & 0.958 $ \pm $ 0.045 & 0.777 $ \pm $ 0.641 \\
	    1-2-3-4 & 18    & 131.15B & 0.30M & 9.32 GB & 0.959 $ \pm $ 0.043 & 0.771 $ \pm $ 0.638 \\
	    1-2-3-4 & 24    & 161.91B & 0.35M & 10.35 GB & 0.961 $ \pm $ 0.047 & 0.765 $ \pm $ 0.640 \\
	    2-2-2-2 & 12    & 103.69B & 0.25M & 8.39 GB & 0.940 $ \pm $ 0.052{*} & 0.876 $ \pm $ 0.705 \\
	    1-2-4-8 & 12    & 103.69B & 0.25M & 8.39 GB & 0.945 $ \pm $ 0.049{*} & 0.848 $ \pm $ 0.719 \\
	    \hline
	    \hline
	    \multicolumn{7}{@{}l}{* indicates a significant difference $(p < 0.05)$ compared with the PLS-Net (2nd row).}
	    \end{tabular}%
		\vspace{-5mm} 
	\end{table}%

	\section{Discussion \& Conclusion}

	In this work, we have presented PLS-Net, a novel deep 3D FCN architecture for automatic pulmonary lobe segmentation in CT images. The PLS-Net has three key features: it adopts 3D DS convolutions in place of regular 3D convolutions for reducing the network's parameters and computational cost, exploits DRDBs to adaptively extract and aggregate multi-scale features, and leverages IR to compensate information loss. Compared to existing 3D FCNs, the efficient architecture and implementation of the PLS-Net along with mixed-precision representations allow it to use less memory to accommodate larger input volumes and more network levels. Consequently, the PLS-Net can learn efficiently from whole high-resolution 3D CT image inputs and outputs on a GPU with 11 GB of memory. This allows high-resolution and global information to be learned for effective lung lobe segmentation. Extensive experiments on a challenging multi-centre dataset demonstrated the effectiveness of the PLS-Net and each of its components. For example, we have shown that 3D DS convolutions can be used as a drop-in replacement for regular 3D convolutions, yielding a 3D FCN that is smaller and computationally cheaper to train and run with negligible loss of accuracy.
	
	To evaluate the performance of our proposed method, we compared it with state-of-the-art approaches evaluated on the same dataset: 210 healthy and pathological lung CT scans acquired with different scanners, scanning protocols and reconstruction parameters. Robustness to such diversity is usually a challenge for existing methods. Experimentally, our PLS-Net showed higher accuracy and better robustness over the compared methods, dealing with challenging cases containing severe pathologies, accessory fissures, and incomplete fissures. The performance gains can be attributed to the deeper high-resolution multi-scale 3D features learnt in the PLS-Net. The PLS-Net also has better efficiency than other FCN-based methods, requiring fewer parameters and computations to achieve the best performance. It can segment the lung lobes from high-resolution 3D CT images in a single forward pass of the network, with an average inference time of 0.8 s on a GTX 1080 Ti GPU, facilitating its adoption in various clinical and research settings.
	
	\begin{figure}[!t]
		\setlength\abovecaptionskip{-0.3\baselineskip}
		\centering
		\includegraphics[width=\linewidth]{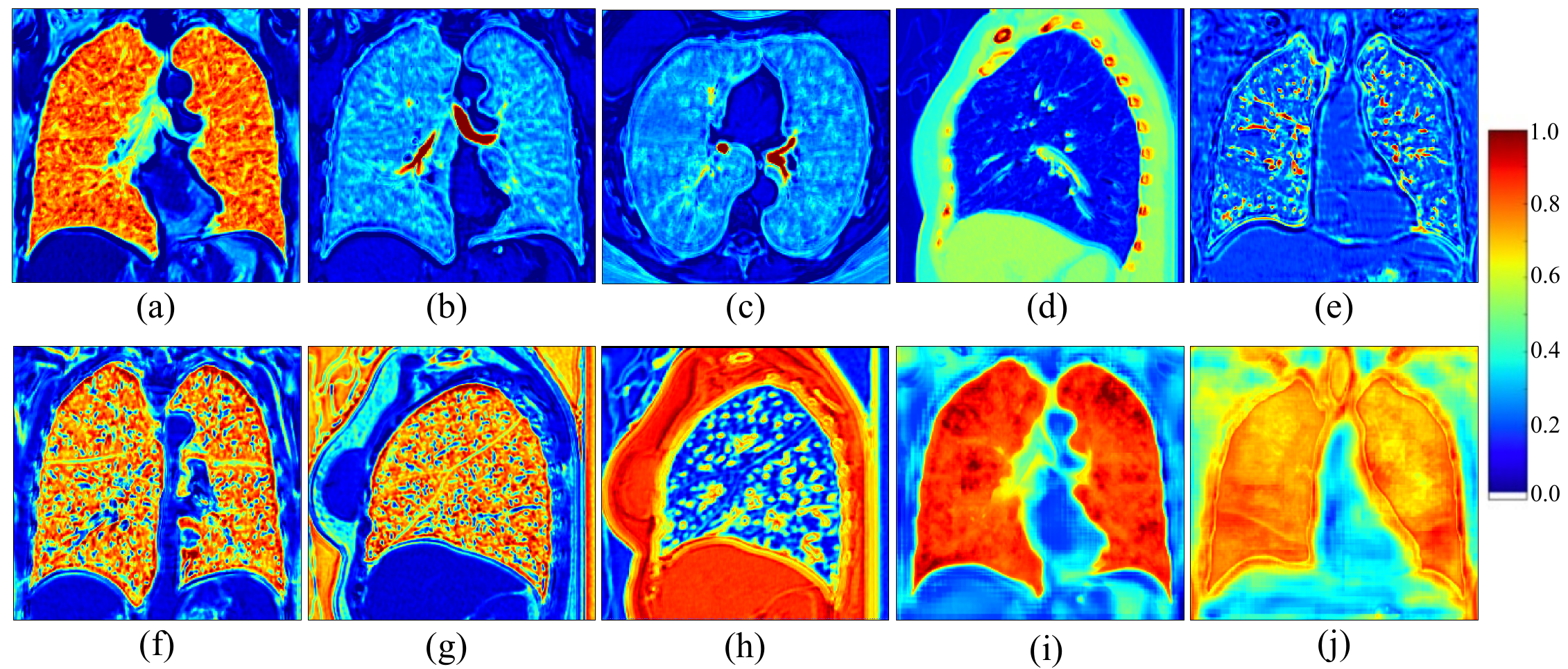}
		\caption{Extracted feature maps from the convolutional layers in the resolution level $l=1$ for the case 100040 in the LTRC dataset. We can see that the network learns to separate the lungs into different compartments (f and g), and identify the rib cage (d) and pulmonary structures: the lungs (a and i), airways (b and c), blood vessels (e and h) and fissures (h and j).}
		\label{fig: feature_maps}
		\vspace{-5mm}
	\end{figure}
	
	Analysing the learned features of a convolutional network could potentially offer new insights and facilitate further research. We have explored what patterns have been automatically learned by the PLS-Net for segmenting the lung lobes. In Figure \ref{fig: feature_maps}, we visualise the 3D feature representations produced by the network when processing a case from the LTRC dataset. Many appearing patterns are hard to interpret, particularly in the deeper layers. Some feature maps in Figure \ref{fig: feature_maps} (displayed in 2D slices for illustration purposes) have an intuitive explanation. It can be observed that the PLS-Net implicitly learns to identify the lungs, fissures, airways and vessels. This indicates that incorporating information from pulmonary structures is beneficial for lobe segmentation. This is similar to the way a human expert performs manual delineation and in line with findings in the literature, where the performance of classical segmentation approaches was improved by combining the pulmonary information \cite{Doel2015, Lassen2013}. Interestingly, the PLS-Net also learns to detect the rib cage and spine, revealing that information on the neighbouring structures of the lungs may provide additional guidance for lobar segmentation.\looseness=-1
	
	There are several limitations in this study. Although the proposed PLS-Net can segment each of the lung lobes more accurately than state-of-the-art approaches, it appears that the same level of accuracy is difficult to obtain for the right middle lobe, which is a consistent trend among existing methods \cite{Doel2012, Lassen2013, Bragman2017, Giuliani2018, George2017, Imran2018, Park2019, Ferreira2018}. It is therefore necessary to evaluate its performance relative to the expert variability. Nevertheless, the dataset used in the study has only one expert annotation per CT scan and does not provide any intra- or inter-observer error margin. Moreover, the segmentation evaluation measures the segmentation fidelity with the expert reference in terms of DSC and ASD, but not the clinical utility of the resulting segmentations for deriving lobar measurements; future work will determine whether the PLS-Net is accurate enough to perform automated lobar volumetry and densitometry for quantitative lobar analysis.

\ifCLASSOPTIONcaptionsoff
  \newpage
\fi



%
\renewcommand*{\bibfont}{\footnotesize}
\bibliographystyle{IEEEtranN}
\bibliography{References}

\end{document}